\documentclass[apj]{emulateapj}
\usepackage{mathtools}
\usepackage{natbib}
\usepackage[breaklinks,colorlinks,
   urlcolor=blue,citecolor=blue,linkcolor=blue]{hyperref}
\usepackage{color}
\usepackage{enumitem}
\usepackage{graphicx,xspace,color,longtable}
\usepackage{rotate,amssymb,amsmath,shadow,bezier,curves,rotating}
\usepackage[toc,titletoc]{appendix}
\usepackage[flushleft]{threeparttable}
\shorttitle{GalWeight Application: A catalog of parameters of 1,800 Galaxy Clusters ($\mathtt{GalWCat19}$)}
\shortauthors{Abdullah et al. 2018}

\begin{document}

\title{GalWeight Application: A publicly-available catalog of dynamical parameters of 1,800 galaxy clusters from SDSS-DR13, ($\mathtt{GalWCat19}$)}

\author{Mohamed H. Abdullah}
\affil{Department of Physics and Astronomy, University of California Riverside, 900 University Avenue, Riverside, CA 92521, USA\\
       Department of Astronomy, National Research Institute of Astronomy and Geophysics, Helwan, 11421 Egypt}
\email{melha004@ucr.edu}

\author{Gillian Wilson}
\affil{Department of Physics and Astronomy, University of California Riverside, 900 University Avenue, Riverside, CA 92521, USA}
%\email{gillian.wilson@ucr.edu}

\author{Anatoly Klypin}
\affil{Astronomy Department, New Mexico State University, Las Cruces, NM 88001, USA
\\
Department of Astronomy, University  of Virginia, Charlottesville, VA 22904, USA}

%\email{aklypin@nmsu.edu}

\author{Lyndsay Old}
\affil{European Space Agency (ESA), European Space Astronomy Centre (ESAC), E-28691 Villanueva de la Ca\~{n}ada, Madrid, Spain\\
Department of Astronomy \& Astrophysics, University of Toronto, Toronto, Canada}

\author{Elizabeth Praton}
\affil{Department of Physics and Astronomy, Franklin \& Marshall College, USA}

\author{Gamal B. Ali}
\affil{Deparment of Astronomy, National Research Institute of Astronomy and Geophysics, Helwan, 11421 Egypt}
%\email{aklypin@nmsu.edu}

\begin{abstract}
Utilizing the SDSS-DR13 spectroscopic dataset, we create a new publicly-available catalog of 1,800 galaxy clusters (GalWeight cluster catalog, $\mathtt{GalWCat19}$) and a corresponding catalog of 34,471 identified member galaxies. The clusters are identified from overdensities in redshift-phase space. The GalWeight technique introduced in Abdullah, Wilson and Klypin (AWK18) is then applied to identify cluster members. The completeness of the cluster catalog ($\mathtt{GalWCat19}$) and the procedure followed to determine cluster mass are tested on the Bolshoi N-body simulations. The 1,800 $\mathtt{GalWCat19}$ clusters range in redshift between $0.01 - 0.2$ and in mass between $(0.4 - 14) \times 10^{14}h^{-1}M_{\odot}$. The cluster catalog provides a large number of cluster parameters including sky position, redshift, membership, velocity dispersion, and mass at overdensities $\Delta = 500, 200, 100, 5.5$. The 34,471 member galaxies are identified within the radius at which the density is 200 times the critical density of the Universe. The galaxy catalog provides the coordinates of each galaxy and the ID of the cluster that the galaxy belongs to. The cluster velocity dispersion scales with mass as  $\log(\sigma_{200})=\log(946\pm52~ \mbox{km} ~ \mbox{s}^{-1}) +(0.349\pm0.142)\log\left[h(z) ~ M_{200}/10^{15}M_\odot\right]$ with scatter of $\delta_{\log\sigma} = 0.06\pm0.04$.
The catalogs are publicly available at the following website\footnote{\url{https://mohamed-elhashash-94.webself.net/galwcat/}}.
\end{abstract}

%_____________________________________________________________________________________________________________________________________
\section{Introduction}

Galaxy clusters are the most massive bound systems in the universe and are uniquely powerful cosmological probes. Cluster dynamical parameters, such as line-of-sight velocity dispersion, optical richness, and mass are closely tied to the formation and evolution of large-scale structures \citep{Bahcall88,Postman92,Carlberg96, Sereno12}. Catalogs of galaxy clusters provide an unlimited data source for a wide range of astrophysical and cosmological applications. In particular, the statistical study of the abundance of galaxy clusters as a function of mass and redshift \citep{Wang98,Haiman01,Reiprich02,Battye03,Dahle06,Lima07,Wen10} is a powerful tool for constraining the cosmological parameters, specifically the normalization of the power spectrum $\sigma_8$ and the matter density parameter $\Omega_m$. Catalogs of galaxy clusters are also interesting laboratories to investigate galaxy evolution under the influence of extreme environments \citep{Butcher78,Dressler80,Goto03,Leauthaud12,Bayliss16,Foltz18}. Moreover, they can be utilized to study  the galaxy-halo connection which correlates galaxy growth with halo growth (e.g., \citealp{Wechsler18}).

Galaxy clusters can be detected based on a number of different properties, such as X-ray emission from hot intracluster gas (e.g., \citealp{Sarazin88,Reichardt13}), the Sunyaev-Zeldovich (SZ) effect \citep{Planck11}, optical (e.g., \citealp{Abell89,denHartog96,Abdullah11}) and infrared emissions (e.g., \citealp{Genzel02,Muzzin09,Wilson09,Wylezalek14}) from stars in cluster members, Stellar Bump Sequence \citep{Muzzin13b}, and the gravitational lensing (e.g., \citealp{Metzler99,Kubo09}). Using current capabilities, both X-ray emission and SZ effect are detectable only for the very deep gravitational potential wells of the most massive systems. They cannot be used to detect the outskirts of massive clusters, or intermediate/low-mass clusters. Thus, current optical surveys of galaxies, such as SDSS, and upcoming surveys such as Euclid \citep{Amendola13}, and LSST \citep{LSST09} are required in order to produce the largest and most complete cluster sample.

Among the most popular applications of galaxy cluster catalogs are scaling relations. Scaling relations of clusters provide insight into the nature of cluster assembly and how the implementation of baryonic physics in simulations affects such relations. Studying these relations for local clusters is also crucial for high-z cluster studies to constrain dark energy (e.g., \citealp{Majumdar04}). Cluster mass is not a directly observable quantity. It can be calculated in several ways such as, the caustic technique \citep{Diaferio99}, the projected mass estimator (e.g, \citealp{Bahcall81}), the virial mass estimator (e.g., \citealp{Binney87}), weak gravitational lensing \citep{Wilson96,Holhjem09}, and application of Jeans equation for the gas density calculated from the x-ray analysis of galaxy cluster \citep{Sarazin88}. However, these methods are observationally expensive to perform, requiring high quality datasets, and are biased due to the assumptions that have to be made (e.g. spherical symmetry, hydrostatic equilibrium, and galaxies as tracers of the underlying mass distribution). Fortunately, the cluster mass can be still indirectly inferred from other observables, the so-called mass proxies, which scale tightly with cluster mass. Among these mass proxies are X-ray luminosity, temperature, the product of X-ray temperature and gas mass (e.g. \citealp{Vikhlinin09,Pratt09,Mantz16a}), optical luminosity or richness (e.g. \citealp{Yee03,Simet17}), and the velocity dispersion of member galaxies (e.g. \citealp{Biviano06,Bocquet15}).

There are many cluster finding methods which rely on optical surveys. For instance, the friends-of-friends (FoF) algorithm is the most frequently usable means for identifying groups and clusters in galaxy redshift data \citep{Turner76,Press82}. It uses galaxy distances derived from spectroscopic or photometric redshifts as the main basis of grouping. Another group of cluster finding methods are halo-based group finders \citep{Yang05,Yang07,Duarte15}. These methods assume some criteria to identify galaxies which belong to the same dark matter halo. An additional cluster finding method is the red-sequence technique, which relies on galaxy colors (e.g., \citealp{Gladders05,Rykoff14}). This red-sequence-based technique assumes the existence of a tight red sequence for clusters, and uses only quiescent galaxies as a proxy of their host cluster environment.  There are other cluster finding methods which are used in the literature, including density-field based methods (e.g., \citealp{Miller05}), matched filter techniques (e.g., \citealp{Kepner99,Milkeraitis10,Bellagamba18}), and the Voronoi-Delaunay method (e.g., \citealp{Ramella01,Pereira17,Soares-Santos11}). These methods are capable of identifying clusters and groups of different richness ranging from a pair of galaxies to very massive clusters with hundreds of galaxies for entire surveys. However, they assume certain criteria and apply fast-run codes to construct catalogs of entire surveys.  This may lead to inaccurate results for recovering the true cluster members because the proposed criteria could be suitable for only some individual clusters depending on their masses and/or dynamical status. Also, most of these methods use photometric redshift to extract cluster catalogs, leading to substantially more uncertainty in cluster membership in comparison to spectroscopically produced catalogs.

It is well-known that galaxy clusters manifest the Finger-of-God effect (see \citealp{Jackson72,Kaiser87,Abdullah13}). This is the distortion of line-of-sight velocities of galaxies both in viral and infall regions due to the cluster potential well, i.e. galaxies peculiar motions. We introduce a simple algorithm, called FG, that identifies locations of clusters by looking for the Finger-of-God effect (FOG). Similar algorithms were introduced in the literature to identity FOG (e.g., \citealp{Yoon08,Wen09,Tempel18}). In this paper, we aim to construct a sample of galaxy clusters using the FG identification in the optical band using a high-quality spectroscopic dataset.  In a previous work (\citealt{Abdullah18}, hereafter AWK18) we introduced a new technique (GalWeight) to assign cluster membership. Galaxy clusters in this catalog are studied individually after assigning galaxy members using the GalWeight technique. 

The paper introduces a catalog of 1,800 galaxy clusters (hereafter, $\mathtt{GalWCat19}$) identified from the spectroscopic dataset of the Sloan Digital Sky Survey-Data Release 13 (hereafter, SDSS-DR13\footnote{\url{https://http://www.sdss.org/dr13}}, \citealp{Albareti17}). We also provide a catalog of 34,471 cluster members. The paper is organized as follows. The data, the FG cluster finding algorithm, and membership identification using GalWeight are introduced in \S \ref{sec:data}.  In \S \ref{sec:dyn} we describe our procedure for calculating the dynamical parameters of each galaxy cluster. Testing the completeness of the catalog and the recovery of dynamical mass using simulations are discussed in \S \ref{sec:sims}. In \S  \ref{sec:results} we describe the $\mathtt{GalWCat19}$ catalog and compare it with some previous catalogs, and introduce the velocity dispersion-mass relation. We summarize our conclusions and future work in  \S \ref{sec:conc}. Throughout the paper we adopt $\Lambda$CDM with $\Omega_m=0.3$, $\Omega_\Lambda=0.7$, and $H_0=100$ $h$ km s$^{-1}$ Mpc$^{-1}$.
%2________________________________________________________________________________________________________________________
\section{Data and clusters identification} \label{sec:data}

\subsection{SDSS sample} \label{sec:SDSS}
Using photometric and spectroscopic database from SDSS-DR13, we extract data for 704,200 galaxies. These galaxies fulfill the following set of criteria: spectroscopic detection, photometric and spectroscopic classification as a galaxy (by the automatic pipeline), spectroscopic redshift between 0.001 and 0.2 (with a redshift completeness $> 0.7$, \citealp{Yang07,Tempel14}), r-band magnitude (reddening-corrected) $< 18$, and the flag SpecObj.zWarning is zero for well-measured redshift. We downloaded the following parameters for each galaxy: photometric object ID, equatorial coordinates (right ascension $\alpha$, declination $\delta$), spectroscopic redshift (z), Petrosian magnitudes in the u, g, r, i and $z$ bands, uncertainties, and extinction values based on \citet{Schlegel98}.

\subsection{Identification of a galaxy cluster} \label{sec:iden}

Galaxy clusters exhibit overdensity regions of $\sim$2-3 orders of magnitude above the background density. One key signature of a galaxy cluster is the distortion of the peculiar velocities of its core members (within $\sim$ 0.5 Mpc from the cluster center) along the line-of-sight. This distortion of FOG appears clearly in a line-of-sight velocity ($v_z$) versus projected radius ($R_p$) phase-space diagram. Here $R_p$ is the projected radius from the cluster center. While, $v_z$ is the line-of-sight velocity of a galaxy in the cluster frame, calculated as $ v_z = ( v_{obs} - v_c)/(1+z_c)$, where $v_{obs}$ is the observed spectroscopic velocity of the galaxy and $z_c$ and $v_c$ are the cluster redshift and velocity, respectively. The observed spectroscopic velocity is calculated as $v_{obs}=c[(z+1)^2-1]/[(z+1)^2+1]$ (relativistic correction). The term $(1+z_c)$ is a correction due to the global Hubble expansion \citep{Danese80} and c is the speed of light. Consequently, the procedure that we follow in this investigation depends on looking for the FOG effect as described below.

\begin{enumerate}

\item We calculate the number density $\rho_{cy}$ of all galaxies within a cylinder of radius $R_{cy} = 0.5 h^{-1}$ Mpc ($\sim$ the width of FOG), and height $3000$ km s$^{-1}$ ($\sim$ the length of FOG) centered on a galaxy i. Note that the radius of the cylinder is equivalent to angular radius $\sin(\theta_{cy})= R_{cy}/D_{c,g}$, where the comoving distance of the galaxy $D_{c,g}$ is calculated as 

\begin{equation} \label{COD}
D_{c,g}=\frac{c}{H_0} \int_0^z \frac{dz'}{\sqrt{\Omega_m(1+z)^3+\Omega_k((1+z)^2+\Omega_\Lambda}}
\end{equation}

\item We sort all galaxies descending from highest to lowest number density with the condition that the cylinder has at least eight galaxies. This means we are aiming to detect all clusters that have at least eight galaxies within a projected distance $R_{p} = 0.5 h^{-1}$ Mpc and velocity range $=\pm1500$ km s$^{-1}$  from the cluster center. The completeness of the catalog is tested on an N-body simulation as described in \S \ref{sec:comp}.

\item Starting with the galaxy with highest number density, we apply the binary tree algorithm (e.g., \citealp{Serra11}) to accurately determine a cluster center ($\alpha_{c}$, $\delta_{c}, z_c$) and a phase-space diagram.

\item We apply the GalWeight technique (see \S \ref{sec:mem}) to galaxies in the phase-space diagram out to maximum projected radius of $R_{p,max} = 10$ $h^{-1}$ Mpc and a maximum line-of-sight velocity of $|v_{z,max}| = 3000$~km~$\mbox{s}^{-1}$ to identify those galaxies within the optimal contour line (see \S \ref{sec:mem} and AWK18). These values are chosen to be sufficiently large to exceed both the turnaround radius (defined in \S\ref{sec:mem}) and the length of the FOG which is typically $\sim7-8~h^{-1}$ Mpc and $\sim 6000$ km s$^{-1}$, respectively, for massive clusters. 

\item Next, using all galaxies enclosed by the optimal contour line (see \S \ref{sec:mem}), we determine the dynamical parameters of each cluster in the catalog (see \S \ref{sec:dyn}).

\end{enumerate}

\subsection{Membership identification: GALWEIGHT} \label{sec:mem}
In AWK18, we introduced GalWeight, a new technique for assigning galaxy cluster membership. AWK18 showed that GalWeight could be applied both to massive galaxy clusters and poor galaxy groups. They also showed that it is effective in identifying members both in the virial and infall regions with high efficiency.

The GalWeight technique works by assigning a weight to a galaxy $i$ according to its position ($R_{p,i}$,$v_{z,i}$) in phase-space diagram. This weight is the product of two separate two-dimensional weights which we refer to as the {\bf{dynamical}} and {\bf{phase-space}} weights:

\noindent 1. The dynamical weight is calculated from the surface number density $\Sigma(R_p)$, velocity dispersion $\sigma_{v_z}(R_p)$, and standard deviation $\sigma_{R_p}(v_z)$ profiles of the cluster as follows. We introduce the function

  \begin{equation} \label{eq:ProbR}
   \mathcal{D}_{R_p} (R_p) = \frac{\Sigma(R_p) \sigma_{v_z}(R_p)}{R_p^\nu},
  \end{equation}
\noindent with the normalization 
  \begin{equation} \label{eq:ProbRN}
   \mathcal{N}_{R_p} = \int_{0}^{R_{p,max}} \mathcal{D}_{R_p} (R_p) dR_p,
  \end{equation}
  
\noindent where $R_{p,max}$ is the maximum projected radius in phase-space and $\nu$ is a free parameter in the range $-1 \lesssim \nu \lesssim 1$ which is introduced to adjust the effect of the distortion of FOG in the core and the distortion of the random motion in the outer region. It is defined as  $\nu =\frac{\sigma_{FOG}(R\leq0.25)}{\sigma_{rand}(0.25<R\leq4))} -1$, where $\sigma_{FOG}$ is the velocity dispersion of the core galaxies and $\sigma_{rand}$ is the velocity dispersion of the galaxies outside the core. Then, Equation \ref{eq:ProbR} is fitted with the following analytical function
  \begin{equation} \label{eq:king}
   \mathcal{W}_{R_p}(R_p)= \mathcal{A}_0\left(1+\frac{R_p^2}{a^2}\right)^{\gamma}+\mathcal{A}_{bg},
  \end{equation}
\noindent  where $a$ is a scale radius ($0 < a \lesssim 1$), $\gamma$ is a slope of the power law ($-2 \lesssim \gamma < 0$), and $\mathcal{A}_{0}$ and $\mathcal{A}_{bg}$ are the central and background weights along the $R_p$-direction.  Also, we define the function

   \begin{equation} \label{eq:ProbV}
   \mathcal{D}_{v_z} (v_z) = \sigma_{R_p} (v_z),
   \end{equation}
\noindent with the normalization 
  \begin{equation} \label{eq:ProbVN}
   \mathcal{N}_{v_z} = \int_{-v_{z,max}}^{v_{z,max}} \mathcal{D}_{v_z} (v_z) dv_z,
  \end{equation}
\noindent where $v_{z,max}$ is the maximum line-of-sight velocity of phase-space. Then, Equation \ref{eq:ProbV} is fitted with the following exponential model
  \begin{equation} \label{eq:Exp}
  \mathcal{W}_{v_z} (v_z) = \mathcal{B}_0 \exp{(b \ v_z)}+\mathcal{B}_{bg},
  \end{equation}
\noindent where $\mathcal{B}_0$ is the central weight, $\mathcal{B}_{bg}$ is the background weight along $v_{z}$ and $b$ is scale parameter ($-0.01 \lesssim b < 0$). Then, the two-dimensional dynamical weight is calculated as 
  \begin{equation} \label{eq:ProbDy}
   \mathcal{W}_{dy}(R_p,v_z)  = \mathcal{W}_{R_p} (R_p) \mathcal{W}_{v_z} (v_z),
  \end{equation}

\noindent 2. The phase-space weight is calculated from the two-dimensional adaptive kernel method that estimates the probability density underlying the data and consequently identifies clumps and substructures in the phase-space \citep{Silverman86,Pisani96}.

The total weight is then calculated as the product of the dynamical and phase-space weights

  \begin{equation} \label{eq:ProbTot}
   \mathcal{W}_{tot}(R_p,v_z)  = \mathcal{W}_{dy} (R_p,v_z) \mathcal{W}_{ph} (R_p,v_z),
  \end{equation}

The optimal total weight value (the optimal contour line) is determined by utilizing the Number Density Method \citep{Abdullah13} in order to separate members and interlopers. Then, we calculate the virial  radius $r_v$ (which is the boundary of the virialized region)  and the turnaround radius  $r_t$ (which is the boundary of the cluster infall region) using the virial mass and NFW mass estimators (\S \ref{sec:dyn}). Finally, the cluster membership are those enclosed by the optimal contour line and within the turnaround radius. The viral radius $r_v$ is the radius within which the cluster is in hydrostatic equilibrium. It is approximately equal to the radius at which the density $\rho=\Delta_{200}\rho_c$, where $\rho_c$ is the critical density of the Universe and $\Delta_{200} = 200$ (e.g., \citealp{Carlberg97}). Therefore, we assume here that $r_v = r_{200}$. The turnaround radius $r_t$ is the radius at which a galaxy's peculiar velocity ($v_{pec}$) is canceled out by the global Hubble expansion. In other words, it is the radius at which the infall velocity vanishes ($v_{inf} = v_{pec} - H~ r = 0$), which can be calculated as the radius at which $\rho = 5.55 \rho_c$  (e.g., \citealp{Nagamine03,Busha05,Dunner06}).

%3_____________________________________________________________________________________________________________________________
\section{Dynamics of galaxy clusters} \label{sec:dyn}
For each cluster, we calculate dynamical parameters i.e., mass, virial and turnaround radii, velocity dispersion, number of spectroscopic members, and concentration as described below.

The cluster mass is estimated from the virial mass estimator (e.g., \citealp{Limber60,Binney87,Rines03}) and NFW mass profile \citep{NFW96,NFW97} as follows. The viral mass estimator is given by

\begin{equation} \label{eq:vir16}
M(<r)=\frac{3\pi N \sum_{i}v_{z, i} (<r)^2}{2G\sum_{i\neq j}\frac{1}{R_{ij}}} 
\end{equation}

\noindent where $v_{z,i}$ is the galaxy line-of-sight velocity and $R_{ij}$ is the projected distance between two galaxies. 

If a system extends beyond the virial radius, Equation~(\ref{eq:vir16}) will overestimate the mass due to external pressure from matter outside the virialized region \citep{The86,Carlberg97,Girardi98}. The corrected virial mass is determined using the following expression:

\begin{equation} \label{eq:vir17}
M_{v}(<r)=M(<r)[1-S(r)], 
\end{equation}

\noindent where $S(r)$ is a term introduced to correct for surface pressure. For an NFW density profile and for isotropic orbits (i.e. the projected, $\sigma_v$, and angular, $\sigma_\theta$, velocity dispersion components of a galaxy in the cluster frame are the same, or equivalently the anisotropy parameter $\beta = 1- \frac{\sigma_\theta^2}{\sigma_r^2} = 0$), $S(r)$ is calculated by

\begin{equation}\label{eq:vir_25}
S(r)=\left(\frac{x}{1+x}\right)^2\left[\ln(1+x)-\frac{x}{1+x}\right]^{-1}\left[\frac{\sigma_v(r)}{\sigma(<r)}\right]^2,
\end{equation}

\noindent where $x=r/r_s$, $r_s$ is the scale radius, $\sigma(<r)$ is the integrated three-dimensional velocity dispersion within $r$, and $\sigma_v(r)$ is a projected velocity dispersion (e.g., \citealp{Koranyi00,Abdullah11}).

The mass density within a sphere of radius $r$ introduced by NFW is given by

  \begin{equation} \label{eq:NFW1}
  \rho(r)=\frac{\rho_s}{x\left(1+x\right)^2}, \end{equation}

\noindent and its corresponding mass is given by

   \begin{equation} \label{eq:NFW11}
   M(<r)=\frac{M_s}{\ln(2)-(1/2)}\left[\ln(1+x)-\frac{x}{1+x}\right],
   \end{equation}

\noindent where $M_s=4\pi\rho_s r^3_s [\ln(2)-(1/2)]$ is the mass within $r_s$, $\rho_s = \delta_s \rho_c$ is the characteristic density within $r_s$ and $\delta_s = (\Delta_v/3) c^3 \left[\ln(1+c) - \frac{c}{1+c}\right]^{-1}$, and the concentration $c = r_v/r_s$  (e.g.,  \citealp{NFW97,Rines03,Mamon13}). 

The projected surface number density of galaxies is given by

   \begin{equation} \label{eq:NFW3}
   \Sigma(<R)=2\rho_s r_s f(x) = \frac{N_s}{\ln(2)-(1/2)} f(x),
   \end{equation}

\noindent where $N_s$ is the number of galaxies within $r_s$ that has the same formula as $M_s$, and $f(x)$ is given by (e.g., \citealp{Golse02,Mamon10})

   \begin{equation}
   f(x) = \begin{cases} \frac{1}{x^2-1}  \left[1-\frac{\cosh^{-1} (1/x)}{\sqrt{1-x^2}}\right] \ \ \ \mbox{if} \  x \ < \ 1\\ 
	                                  \frac{1}{3}  \ \ \ \ \ \ \ \  \ \ \ \ \ \ \ \ \ \  \ \ \ \ \ \ \ \ \  \ \ \ \mbox{if} \  x \ = \ 1 \\ 
	                                  \frac{1}{x^2-2} \left[1-\frac{\cos^{-1}    (1/x)}{\sqrt{x^2-1}}\right]\   \ \ \ \mbox{if} \  x \ > \ 1
   \end{cases} \end{equation}

The projected number of galaxies within a cylinder of radius R is given by integrating the NFW profile (Equation~(\ref{eq:NFW1})) along the line of sight (e.g., \citealp{Bartelmann96,Zenteno16})

   \begin{equation} \label{eq:NFW2}
   N(<R)=\frac{N_s}{\ln(2)-(1/2)} g(x),
   \end{equation}

\noindent where $g(x)$ is given by (e.g., \citealp{Golse02,Mamon10})

   \begin{equation}
   g(x) = \begin{cases} \ln(x/2) + \frac{\cosh^{-1} (1/x)}{\sqrt{1-x^2}} \ \ \ \mbox{if} \  x \ < \ 1\\ 
	                      1-\ln(2)  \ \ \ \ \ \ \ \ \ \ \ \ \ \ \ \ \ \  \  \mbox{if} \  x \ = \ 1 \\ 
	                      \ln(x/2) + \frac{\cos^{-1} (1/x)}{\sqrt{x^2-1}}  \ \  \ \ \ \mbox{if} \  x \ > \ 1
   \end{cases} \end{equation}

Given the projected radii of galaxies in each cluster, we fit $r_s$ with a maximum-likelihood estimation (MLE) by finding the value of  $r_s$ that minimizes the probability

   \begin{equation}
   -\ln{L}=-\sum_i \ln{\frac{x_i\Sigma(x_i)}{\int^{x_{max}}_{0}  x_i \Sigma(x_i) dx}}     
   \end{equation}
\noindent where $x_{max} = R_{max}/r_s$ and $R_{max}$ is a maximum projected radius. In practice, we search for the best value of $r_s$ that gives minimum likelihood within $R_{max} \lesssim 3R_{200}$, where $R_{200}$ is initially calculated from the uncorrected virial mass estimator (Equation \ref{eq:vir16}). We determine the uncertainty of  $1\sigma$ confidence interval by $- \ln L =- \ln L_{ML} + 0.5$, where $\ln L_{ML}$ is the maximum likelihood (see e.g., \citealp{Koranyi00,Mamon10,Mamon13}).

To summarize the procedure described above to calculate the corrected virial mass and NFW mass profile for each cluster:  we first fit $r_s$ for each cluster to get $S(r)$ (Equation~\ref{eq:vir_25}); we then calculate the corrected virial mass  $M_v(<r_{200})$ (Equation~\ref{eq:vir17}) at the virial radius $r_{200}$\footnote{Throughout the paper we interchangeably call $r_v$ and $r_{200}$ for the virial radius. In practice, the virial radius at which the cluster is in hydrostatic equilibrium cannot be determined. We follow convention and assume that $r_v$ is at $\rho= 200 \rho_c$.}, at which $\rho= 200 \rho_c$; we then calculate the NFW mass profile from Equation~\ref{eq:NFW11}; finally, we determine the dynamical parameters (radius, number of members, velocity dispersion and mass) at overdensities of $\Delta = 500, 200, 100, 5.5$.
%4____________________________________________________________________________________________________________________________
\section{Application to Simulations}\label{sec:sims} 
In \S \ref{sec:comp} we test the completeness of the FG algorithm (see \S \ref{sec:iden}) using the Bolshoi N-body simulation \citep{Klypin16}. In \S \ref{sec:recover} we test the procedure described in \S \ref{sec:dyn} to recover a cluster mass using two mock catalogs recalled from \citet{Old15}. Note that the efficiency of GalWeight for assigning cluster membership has already been tested on Bolshoi \& MDPL2 N-body simulations, and has  been found to be $> 98\% $ accurate in correctly assigning cluster membership (see Table 1 in AWK18).

\subsection{{Catlaog Completeness as a Function of Cluster Mass and Redshift}} \label{sec:comp}
In this section we investigate the completeness of the FG algorithm to identify locations of clusters with at least eight spectroscopic galaxies (see \S \ref{sec:iden}). In order to achieve this investigation we apply the FG algorithm to the Bolshoi\footnote{\url{https://www.cosmosim.org}} simulation. The Bolshoi simulation is an N-body simulation of $2048^3$ particles in a box of comoving length 250 $h^{-1}$ Mpc, mass resolution of $1.35 \times 10^8$ $h^{-1}$ M$_{\odot}$, and gravitational softening length of 1 $h^{-1}$ kpc (physical) at low redshifts. It was run using the Adaptive Refinement Tree (ART) code \citep{Kravtsov97}. It assumes a flat $\Lambda$CDM cosmology, with cosmological parameters ($\Omega_\Lambda$ = 0.73, $\Omega_m$ = 0.27, $\Omega_b$ = 0.047, $n$ = 0.95, $\sigma_8$ = 0.82, and $h$ = 0.70.  Halos are identified using the Bound Density Maximum (BDM) algorithm \citep{Klypin97,Riebe13}, that was extensively tested (e.g., \citealp{Knebe11}) which identifies local density maxima, determines a spherical cut-off for the halo with overdensity equal to 200 times the critical density of the Universe ($\rho = 200 \rho_{c}$), and removes unbound particles from the halo boundary. Among other parameters, BDM provides a virial masses and radii. The virial mass is defined as $M_{v} =\frac{4}{3} \pi 200 \rho_{c} r_{v}^3$ (see \citealp{Bryan98,Klypin16}). The halo catalogs are complete for halos with circular velocity $v_c \geq 100$ km s$^{-1}$ (e.g., \citealp{Klypin11,Busha11}).

In order to investigate the completeness and purity of FG we construct a light-cone from Bolshoi as follows. We treat all subhalos as galaxies and assume the line-of-sight to be along the z-direction and the projection to be on the x-y plane. We calculate right ascension ($\alpha$), declination ($\delta$) and radial distance ($D_c$) in real-space as, 

\begin{equation} 
\begin{cases} 
D_c= \sqrt{x^2+y^2+z^2}\\
\alpha = \arctan(yx)\\
\delta = \arcsin(x/d),
\end{cases}
\end{equation}
\noindent where x, y, and z are the co-moving coordinates along the principal axes of the simulation box.

The cosmological redshift $z_{cosm}$ of a galaxy is determined by inverting $D_c$, using the distance-redshift relation for the given simulation cosmology (see Equation \ref{COD}). The line-of-sight peculiar redshift in a cluster-frame is calculated as

\begin{equation} 
z_{pec} = \left(\frac{x}{D_c} v_x + \frac{y}{D_c} v_y + \frac{z}{D_c} v_z\right)/ c,
\end{equation} 
\noindent  where $v_x$, $v_y$, and $v_z$ are the peculiar velocity components and c is the speed of light.

Finally the observed redshift is calculated as 

\begin{equation} 
(1+z_{obs}) = (1+z_{cosm}) (1+z_{pec})
\end{equation} 

For Bolshoi, we have about 791 clusters with masses $\geq 0.40\times10^{14} h^{-1} M_{\odot}$. We triple the number of clusters by operating the same task on the other two line-of-sights (x- and y-directions) and the other two projections (x-z, and y-z planes). We apply the FG algorithm to each light-cone. We then match the detected clusters with the true simulated ones within a radius of 1.5 $ h^{-1}$Mpc and velocity gap of $\pm 1500$ km s$^{-1}$ (see \S \ref{sec:match}).

The completeness and purity of FG are defined as (e.g., \citealp{Hao10}) 

\begin{equation} \label{eq:DR1}
\begin{cases} 
\mathcal{C}_{bin} (x)= \frac{N^{match}_{sim}(x_1\leq x<x_2)}{N_{sim}(x_1\leq x<x_2)}\times 100\\

\\

\mathcal{C}_{cum} (x)= \frac{N^{match}_{sim}(x \geq x_1)}{N_{sim}(x \geq x_1)}\times 100 ,
\end{cases}
\end{equation}

\begin{equation} \label{eq:DR2}
\begin{cases} 
\mathcal{P}_{bin} (x)= \frac{N^{match}_{sim}(x_1\leq x<x_2)}{N_{det}(x_1\leq x<x_2)}\times 100\\
\\
\mathcal{P}_{cum} (x)= \frac{N^{match}_{sim}(x \geq x_1)}{N_{det}(x \geq x_1)}\times 100 ,
\end{cases}
\end{equation}

\noindent where  $\mathcal{C}_{bin}$ and $\mathcal{P}_{bin}$ are the completeness and purity between $x_1$ and $x_2$, $\mathcal{C}_{cum}$ and $\mathcal{C}_{cum}$ represent the cumulative rates, and $x$ is a parameter that represents cluster mass or richness (number of galaxies). Here, $N_{det}$ is the total number of clusters detected by FG, $N_{sim}$ is the total number of simulated clusters, and $N^{match}_{sim}$ is the number of clusters which are detected by FG and matched with the simulated clusters.

% Fig1___________________________________________________
\begin{figure*} \hspace*{0.5cm}
\includegraphics[width=21cm]{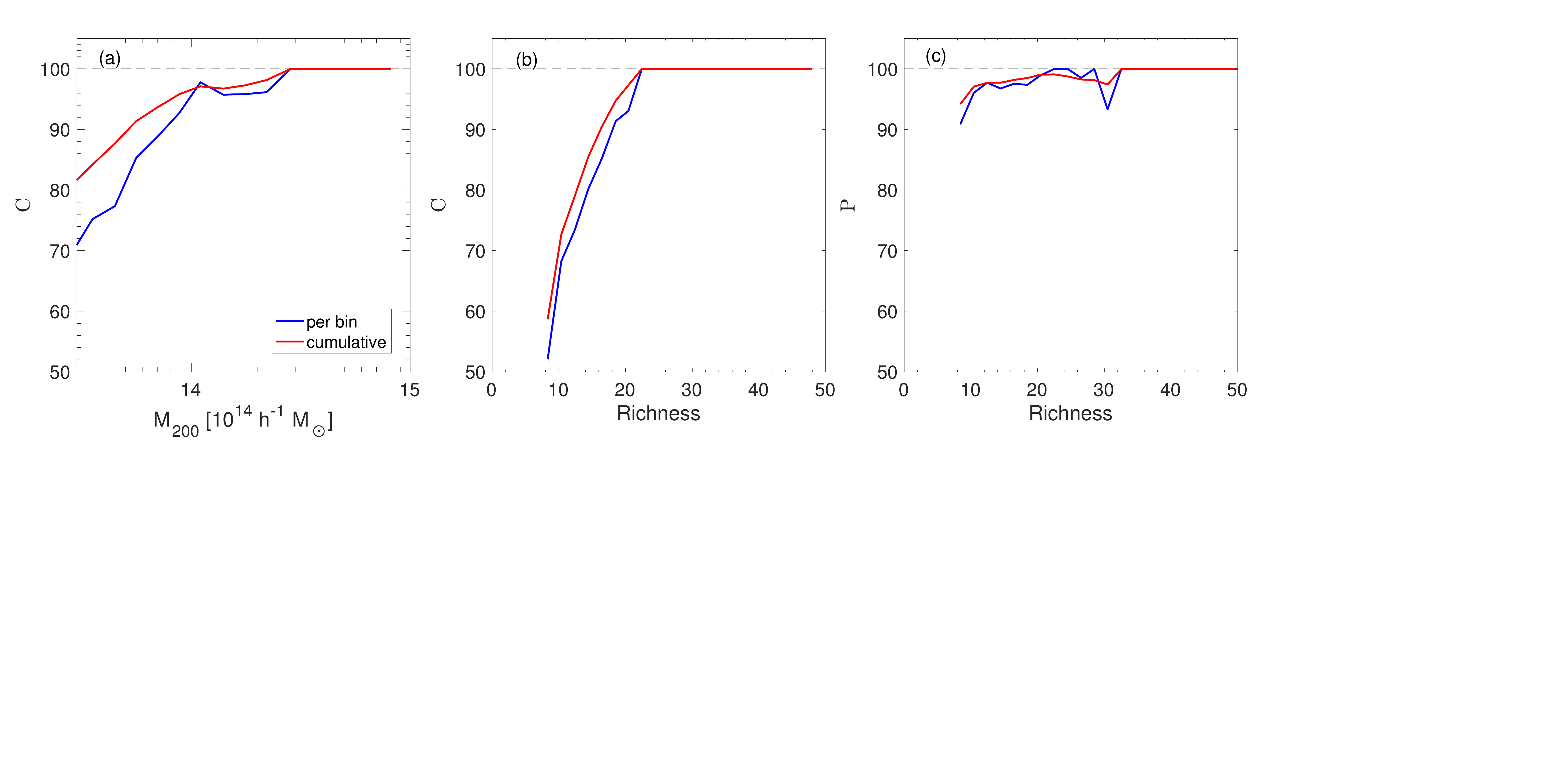} \vspace{-5cm}
\caption{Completeness and purity of the FG algorithm. (a): completeness of FG applied to the Bolshoi clusters as a function of cluster mass for at least eight galaxies in a cylinder of radius $R_{cy} = 0.5~h^{-1}$ Mpc and height $3000$ km s$^{-1}$  (see \S \ref{sec:iden}). (b): completeness of FG as a function of richness (number of galaxies in the cylinder). (c): purity of FG as a function of richness. The blue lines represent the rates per bin, and the red lines represent the cumulative rates. 
%which is  $\sim100\%$ for clusters with masses $M_{200} > 2 \times 10^{14}~h^{-1}M_{\odot}$, and $\sim 85\%$ for clusters with masses $M_{200} > 0.4 \times 10^{14} h^{-1}$ M$_{\odot}$. 
}
\label{fig:comp}
\end{figure*}

% Fig2___________________________________________________
\begin{figure*} \hspace{0.5 cm}
\includegraphics[width=21cm]{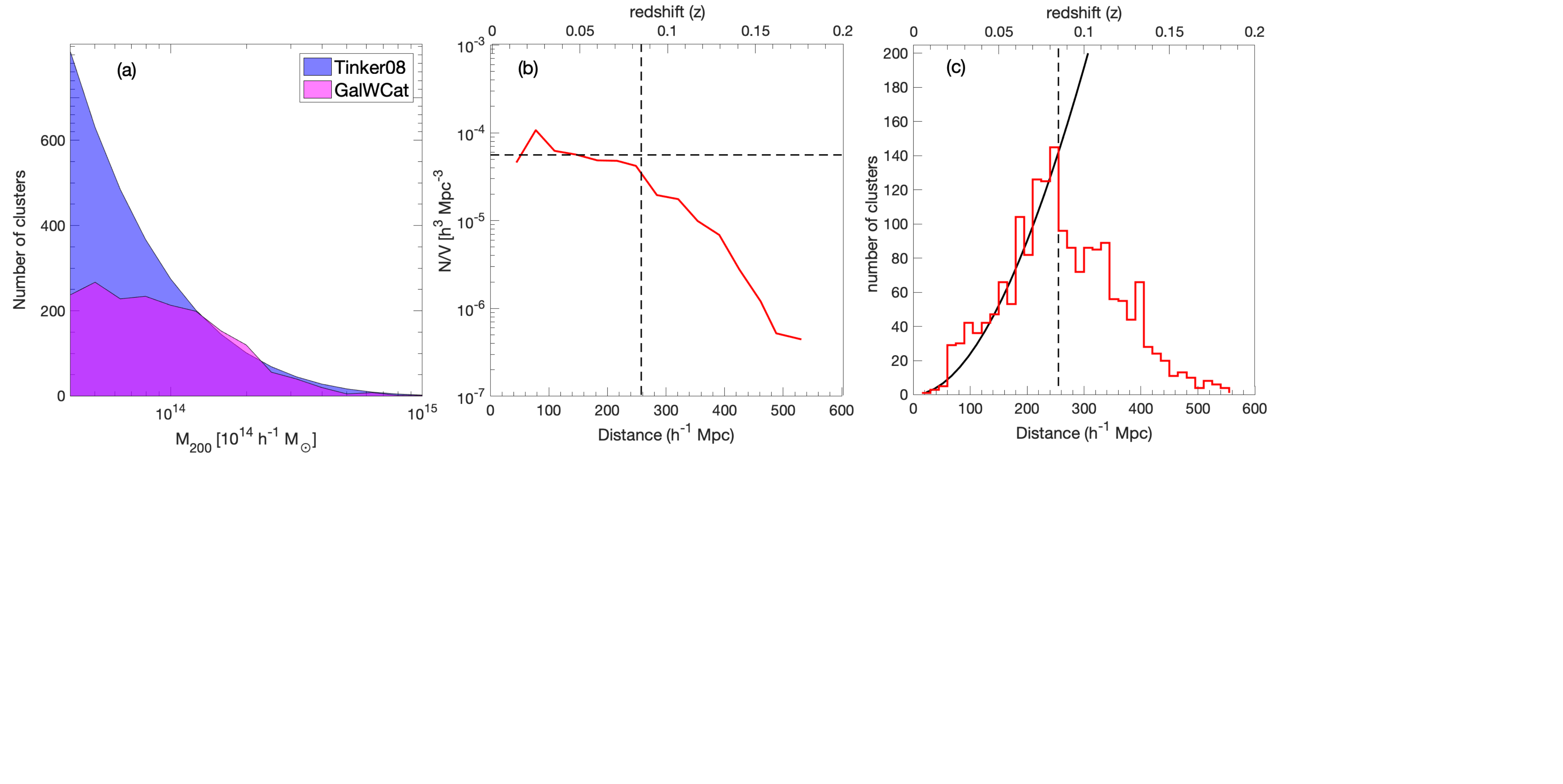} \vspace{-4.75 cm}
\caption{Completeness of GalWeight catalog. (a): the abundance of clusters as a function of mass for $\mathtt{GalWCat19}$ (red area) compared to the abundance of clusters predicted by \citet{Tinker08} model (blue area). (b): cluster number density as a function of comoving distance for $\mathtt{GalWCat19}$. The solid black line shows the number density the sample and the dashed black horizontal line represents the number density of $5.6\times10^{-5}~h^{3}$ Mpc$^{-3}$ averaged for the overall sample within distance $D\leq225~ h^{-1}$. (c): number of clusters as a function of comoving distance. The dashed black line shows the expectation for a completed volume-limited sample with a density of $5.6\times10^{-5}~h^{3}$ Mpc$^{-3}$ for $\Omega_m=0.3$ and $\Omega_\Lambda=0.7$.}

\label{fig:compz}
\end{figure*}

Figure \ref{fig:comp}.a shows the compactness of FG as a function of cluster mass for at least eight galaxies in a cylinder of radius $R_{cy} = 0.5~h^{-1}$ Mpc and height $3000$ km s$^{-1}$  (see \S \ref{sec:iden}). As shown, the cumulative compactness (red line) is  $\sim 100\%$ for clusters with masses $M_{200} > 2 \times 10^{14}~h^{-1}M_{\odot}$, while it drops to $\sim 85\%$ for clusters with masses $M_{200} > 0.4 \times 10^{14}~h^{-1}M_{\odot}$.  Figure \ref{fig:comp}.b presents the compactness of FG as a function of richness (number of galaxies in the cylinder), and Figure \ref{fig:comp}.c shows the purity of FG.

The completeness in mass of the $\mathtt{GalWCat19}$ catalog can be investigated by calculating the abundance of clusters predicted by a theoretical model and compare it with the abundance of $\mathtt{GalWCat19}$ clusters. The halo mass function (HMF), defined as the number of dark matter halos per unit mass per unit comoving volume of the universe, is given by 

\begin{equation}
\label{eq:hmf}
  \frac{dn}{d\ln M} =  f(\sigma) \frac{\rho_0}{M} \left|\frac{d\ln\sigma}{d\ln M}\right|;
\end{equation}

\noindent here $\rho_0$ is the mean density of the universe, $\sigma$ is the rms mass variance on a scale of radius $R$ that contains mass $M = 4 \pi \rho_0 R^3/3$ , and $f(\sigma)$ represents the functional form that defines a particular HMF fit. 

We adopt the functional form of \citet{Tinker08} (hereafter Tinker08) to calculate the HMF and consequently the predicted abundance of clusters. For more detail about the calculation of the HMF we refer the reader to e.g.,  \citet{Press74,Sheth01,Jenkins01,Warren06,Tinker10b,Behroozi13a}. The HMF is calculated using the publicly available \verb|HMFcalc| \footnote{\url{http://hmf.icrar.org/}} code \citep{Murray13b}. We adopt the following cosmological parameters: $\Omega_m = 0.307$, $\Omega_\Lambda = 0.693$, $\sigma_8 = 0.823$, CMB temperature $T_{cmb} = 2.725 K^\circ$, baryonic density $\Omega_b = 0.0486$, and spectral index $n = 0.967$ \citep{Planck14}, at redshift $z = 0.089$ (the mean redshift of $\mathtt{GalWCat19}$). 

Figure \ref{fig:compz}.a shows the abundance of clusters as a function of mass for $\mathtt{GalWCat19}$ (red area) compared to the abundance predicted by Tinker08 (blue area). As shown, the $\mathtt{GalWCat19}$ is complete in mass for $M_{200} \gtrsim 1 \times 10^{14}~h^{-1}M_{\odot}$, while it drops off below this mass.

We also investigate the completeness of $\mathtt{GalWCat19}$ as a function of redshift or comoving distance. The left panel of Figure \ref{fig:compz}.b shows the number density of clusters as a function of comoving distance. The number density is almost constant within comoving distance $\sim 225 h^{-1}$Mpc ($z \sim 0.088$), except for the nearby regions where the cosmic variance due to the small volume has a large effect. The number density drops catastrophically beyond $\sim 225 h^{-1}$Mpc. Figure \ref{fig:compz}.c presents the abundance of clusters as a function of distance. Comparing the data with the expectation of a constant number density (shown as the dashed black line, $5.6\times10^{-5}~h^3$ Mpc$^{−3}$) shows that $\mathtt{GalWCat19}$ is incomplete beyond $\sim 225 h^{-1}$Mpc. The dependence of the number density on both the cluster mass and selection function of $\mathtt{GalWCat19}$ is investigated in detail in Abdullah et al. (2019b, in prep) which studies the cluster mass function.

\subsection{Effectiveness of Cluster Mass Estimation} \label{sec:recover}
In order to test our procedure to determine cluster masses (see \S \ref{sec:dyn}) we use two distinct mock catalogs utilized in \citet{Old15,Old18} to investigate the performance of a variety of cluster mass estimation techniques. These two mock catalogs are derived from the Bolshoi DM simulation. The first mock catalog places galaxies onto the Bolshoi DM simulation by a Halo Occupation Distribution (HOD) model.  The specific model in this case is referred to as HOD2, and is an updated version of the model described in \citet{Skibba06,Skibba09}. The second one depends on the Semi-Analytic Galaxy Evolution (SAGE) galaxy formation model \citep{Croton16}, which is an updated version of that described in \citep{Croton06}. This mock catalog is referred to as SAM2. We refer the reader to \citet{Old14,Old15} for more detail regarding the construction of these catalogs.

\citet{Old15} performed an extensive comparison of 25 galaxy-based cluster mass estimation methods using the HOD2 and SAM2 catalogs. Following \citet{Old15}, we examine the performance of our procedure to recover cluster mass by calculating the root-mean-square (rms) difference between the recovered and input log mass, defined as 

\begin{equation}
rms = \sqrt{\frac{1}{N}\sum_i^N \left(\log M_{i,true}-\log M_{i,rec}\right)^2}
\end{equation}
\noindent where $M_{i,true}$ is the true mass of the cluster and $M_{i,rec}$ is its recovered or estimated mass. 

We also test the performance of the procedure by calculating the scatter in the recovered mass, $\sigma_{M_{rec}}$ (delivers a measure of the intrinsic scatter), the scatter about the true mass, $\sigma_{M_{true}}$, and the bias at the pivot mass, where the pivot mass is taken as the median log mass of the input cluster sample ($\log M_{true} = 14.05$). For these three statistics, we assume a linear relationship between the recovered and true log mass (see section 4.2 in \citealp{Old15} for a full description of these statistics and, e.g., \citealp{Hogg10,Sereno15,Andreon17}).

% Fig3___________________________________________________
\begin{figure*} \hspace*{0.5 cm}
\includegraphics[width=16cm]{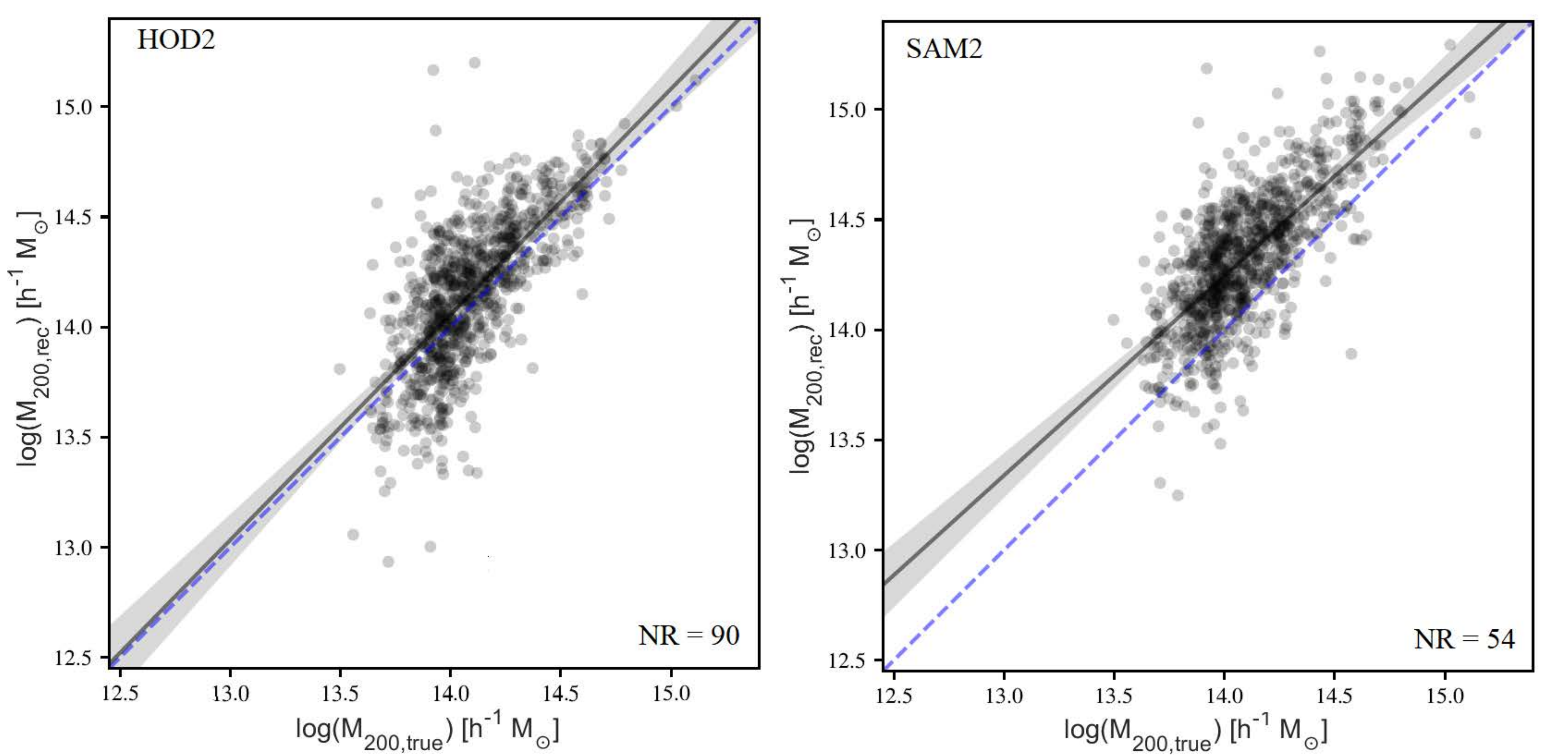} \vspace{0.0 cm}
\caption{Recovered versus true cluster mass applied to the HOD2 (left) and the SAM2 (right) catalogs. The blue dashed lines represent the one-to-one relation. The solid black lines show the linear relationship between the recovered and true log mass. NR in the legend represents the number of missing clusters out of 1000 simulated clusters.}
\label{fig:HODSAM}
\end{figure*}

We apply our procedure (see \S \ref{sec:dyn}) on the HOD2 and SAM2 catalogs to calculate cluster mass. Figure \ref{fig:HODSAM} shows the recovered versus true cluster mass applied to the HOD2 (left) and the SAM2 (right) catalogs (see  Figures 2 and 4 in \citealp{Old15} for comparison).
We find that the procedure performs very well in comparison to all of the other 25 methods and results in lower values of 
the aforementioned statistical quantities than most of these methods for both the HOD2 and SAM2 models. 
Quantitatively, $rms$, $\sigma_{M_{rec}}$, $\sigma_{M_{true}}$, and bias are 0.24, 0.23, 0.23, and 0.06 for HOD2 and 0.32, 0.21, 0.23, and 0.24 for SAM2, respectively. These values are amongst the lowest of all the methods which calculate the cluster mass from the galaxy velocity dispersion except for the bias calculated for SAM2 which returns a slightly higher value (see Table 2 in \citealp{Old15} for comparison). We use two different mock catalogs that have been constructed in an inherently different in the way for the purpose of observing any potential variation in mass estimation technique assessment due to assumptions made in constructing the mock catalogs.

The scatters and bias calculated above have a number of causes. Specifically, factors that introduce scatter when using the virial mass estimator include:
(i) the assumption of hydrostatic equilibrium, projection effect, and possible velocity anisotropies in galaxy orbits, and the assumption that halo mass follows light (or stellar mass); (ii) presence of substructure and/or nearby structure  such as cluster, supercluster, to which the cluster belongs, or filament (see e.g., \citealp{The86,Merritt88,denHartog96,Fadda96,Girardi98,Abdullah13} for more details about these effects); (iii) presence of interlopers in the cluster frame due to the triple-value problem, for which there are some foreground and background interlopers that appear to be part of the cluster body because of the distortion of phase-space \citep{Tonry81,Abdullah13}; (iv) identification of cluster center (e.g., \citealp{Girardi98,Zhang19}).

% Fig4_____________________________________________________________________________________
\begin{figure*} \hspace*{-8.5cm}
\includegraphics[width=34.5cm]{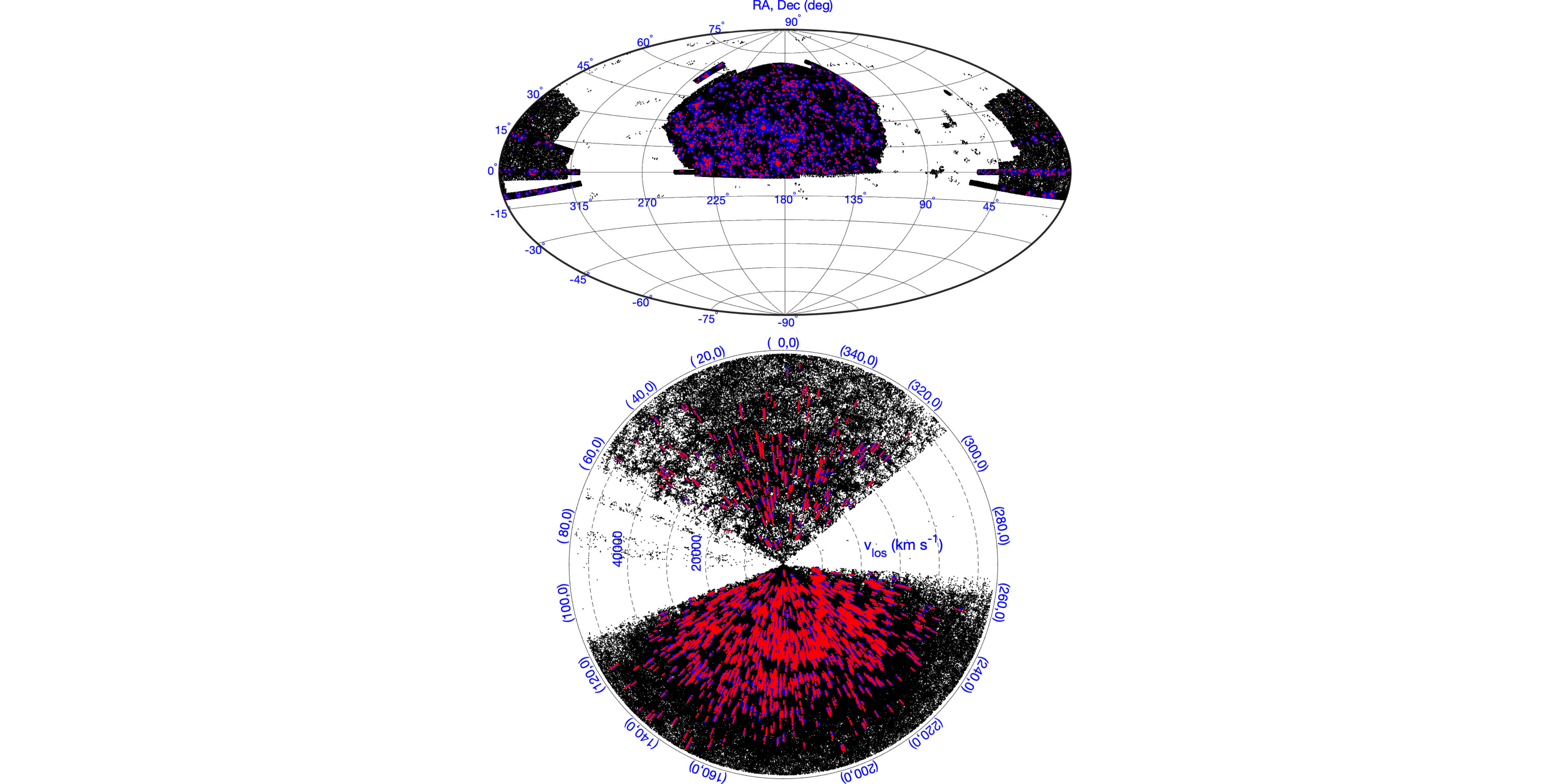} \vspace{-0.cm}
\caption{Top panel: Aitoff projection in celestial coordinates. Bottom panel: light cone diagram.
The black points represent the distribution of all galaxies in the sample, while the blue and red points represent the distribution of 1,800 clusters members identified by GalWeight which are within $r_{200}$ and  $r_{5.5}$, respectively (see \S \ref{sec:dyn}).}
\label{fig:AitoffSlice}
\end{figure*}

%2*******************************************************************************                                                                                       
\begin{table*} \centering
\caption{Coordinates, dynamical parameters at $R_{200}$,  and NFW parameters for the first 15 clusters in the $\mathtt{GalWCat19}$ catalog (see Appendix \ref{app:catalogs}).}
\label{tab:Virial}
\scriptsize
\begin{tabular}{cccccccccc}
\hline
ID &  $\alpha$  & $\delta$ &$z_c$ &  $r_{200}$&$N_{200}$&$\sigma_{200}$&$M_{200}$&$r_s$ &$M_s$\\
&(deg)&(deg)&& ($h^{-1}$ Mpc)&&(km s$^{-1})$&($10^{14}~h^{-1}$M$_{\odot}$)&($h^{-1}$ Mpc)&($10^{14}~h^{-1}$M$_{\odot}$)\\
(1)&(2) &(3)&(4)&(5)&(6)&(7)&(8) & (9)&(10)\\ 
\hline
   01 & 230.7 &  27.74 &  0.0732 & 1.759 & 167 &  $1042^{+100.8}_{-83.23}$ & $13.55\pm 3.569$ & $0.41\pm 0.108 $&  $3.061\pm 0.8063$\\
   02 & 227.6 &   33.5  &  0.1139 & 1.511  &  63  & $926.8^{+118.7}_{-92.04}$ & $8.947\pm 2.349$ & $0.32\pm 0.084 $&  $1.880\pm 0.4935$\\
   03 & 194.9 &  27.91  & 0.0234 & 1.545 & 672 & $932.6^{+55.31}_{-47.76}$& $8.757\pm 2.229$ & $0.33\pm 0.084 $&  $1.852\pm 0.4715$\\
   04 & 258.2 &  64.05 & 0.0810 & 1.453 & 155 & $881.2^{+82.08}_{-70.22}$ & $7.702\pm 2.082$ & $0.37\pm  0.1$ &  $1.865\pm 0.5039$\\
   05 & 209.8 &  27.97 & 0.0751 & 1.449 &  77   & $842.7^{+96.16}_{-83.23}$ & $7.589\pm 2.168$ & $0.14\pm 0.04 $&  $0.966\pm 0.276$\\
   06 & 227.7 &  5.823 & 0.0784 & 1.431 & 128  & $886.5^{+86.03}_{-71.71} $& $7.332\pm 0.4412 $& $1.13\pm 0.068$ &  $5.459\pm 0.3285$\\
   07 & 255.7 &   33.50& 0.0878 & 1.423 &  84  & $887.6^{+130.6}_{-95.31} $& $7.281\pm 2.118   $& $0.22\pm 0.064 $&  $1.229\pm 0.3574$\\
   08 & 231.0 &  29.89 &  0.1138 &  1.390 &  80 & $860.2^{+99.22}_{ -87.6} $& $6.952\pm 1.106  $& $0.93\pm 0.148$ &  $4.265\pm 0.6787$\\
   09 & 239.6 &  27.22 & 0.0898 & 1.379 & 150 & $838.2^{+92.67}_{-77.16} $& $6.632\pm 1.404 $& $0.85\pm 0.18 $&  $3.709\pm 0.7855$\\
   10 & 240.5 &  15.92 &  0.0370 & 1.401 & 299 & $771.3^{+62.73}_{-52.71} $&$ 6.618\pm 1.655  $ & $0.16\pm 0.04 $&  $0.926\pm 0.2315$\\
   11 & 257.4 &  34.47 &  0.0849 & 1.378 &  93  & $846.6^{+117.8}_{-92.73} $& $6.583\pm 2.088$ & $0.29\pm 0.092$ &  $1.377\pm 0.437$\\
   12 & 255.7 &  34.05 &  0.1002 &  1.340&  72  & $833.3^{+99.68}_{-80.46}$ & $6.154\pm 1.783 $& $0.29\pm 0.084 $&  $1.314\pm 0.3807$\\
   13 & 2.939 &  32.42 &  0.1017 & 1.322 &  35  & $861.5^{+159.5}_{-119.3} $& $5.915 \pm 1.577    $& $0.15\pm 0.04 $& $0.8245\pm 0.2199$\\
   14 & 53.59 & -1.166 &  0.1381 & 1.272 &  28   &$ 878.4^{+301.3}_{-176.4} $& $5.462\pm 1.923 $ & $0.25\pm 0.088 $&  $1.087\pm 0.3827$\\
   15 & 358.5 & -10.39 & 0.0766 & 1.284 & 109 &$ 753.3^{+88.64}_{-75.14} $& $5.287\pm  1.41  $ & $0.36\pm 0.096 $&  $1.384\pm 0.3691$\\
\hline
\end{tabular}
\begin{tablenotes}
\item
\noindent
Columns: (1) cluster ID; (2) right ascension; (3)  declination; (4) redshift, (5-8) radius and its corresponding number of members, velocity dispersion and mass at overdensity of $\Delta= 200$; (9-10) scale radius and its corresponding mass of NFW model.
\end{tablenotes}
\end{table*} 

%5_______________________________________________________________________________________________________________
\section{GalWeight cluster catalog, $\mathtt{GalWCat19}$} \label{sec:results}

\subsection{Dynamical Parameters}
As discussed in \S  \ref{sec:iden} we identify the location of a galaxy cluster in a cylinder of radius $R_{cy} = 0.5~h^{-1}$ Mpc and height $3000$ km s$^{-1}$ with the condition that the cylinder has at least eight galaxies. We then apply the GalWeight technique to assign its membership (see \S \ref{sec:mem}). Then, using the virial mass estimator we determine the cluster virial mass assuming that the virial radius is at  $\rho = 200 \rho_c$ (see \S \ref {sec:dyn}). Finally, we select all galaxy clusters of virial mass $M_{200} \geq 0.4\times 10^{14}$ $h^{-1}$ M$_{\odot}$. Following this procedure we get a catalog of 1,800 clusters with virial mass in the range $(0.40 - 14) \times 10^{14}$ $h^{-1}$ M$_{\odot}$ and in a redshift range $0.01 \leq z \leq 0.2$. We refer to this 1,800 galaxy cluster sample as $\mathtt{GalWCat19}$. We exclude overdensity regions (locations of galaxy clusters) for which the FOG effect is indistinct because of interactions between different clusters in these regions.

The distribution of all galaxies in the sample (black points) and the cluster members identified by GalWeight and within $r_v$ (red points) and $r_t$ (blue points) are shown in Figure \ref{fig:AitoffSlice}. The distortion of the line-of-sight velocity or the FOG effect is shown clearly for each cluster. 

As discussed in \S \ref{sec:dyn} we use the virial mass estimator to determine the virial mass at the virial radius $r_{200}$ of each cluster.  Then, using NFW mass profile we determine the dynamical parameters of each cluster at overdensities of $\Delta = [500, 200, 100, 5.5]$. Note that we assume the virial radius is at $\Delta$ = 200 and turnaround radius is at $\Delta$ = 5.5 (see \S \ref {sec:dyn}). The derived parameters for each cluster are radius, number of members, velocity dispersion and mass at each of the different overdensities, plus the NFW parameters: scale radius, mass at scale radius, and concentration $c=r_{200}/r_s$ (see Appendix \ref{app:catalogs}). Table \ref{tab:Virial} shows the coordinates, dynamical parameters at at $R_{200}$, and  NFW parameters for the first 15 clusters in the $\mathtt{GalWCat19}$ catalog. 

The $\mathtt{GalWCat19}$ release consists of two catalogs.  The first catalog is for the coordinates and the dynamical parameters of each galaxy cluster and the second one is for the coordinates of member galaxies belonging to each cluster. The two catalogs are described in Appendix A, and made available in their entirety at the link\footnote{\url{https://mohamed-elhashash-94.webself.net/galwcat}}. The uncertainty of the virial mass estimator is calculated using the limiting fractional uncertainty $\pi^{-1} (2~\ln~N)^{1/2}N^{-1/2}$ \citep{Bahcall81}. Note that throughout the paper the velocity dispersion is calculated using the classical standard deviation $\sigma_v = \left[(n-1)\right]^{-1}\sum_i v_z^2$, where $v_z$ is the line-of-sight velocity of a galaxy in the cluster frame (e.g., \citealp{Munari13,Tempel14,Ruel14}). The uncertainty of the velocity dispersion is calculated via performing bootstrap resampling (with 1000 resamples).
%##################################################################################

\subsection{GalWeight Catalog Matching} \label{sec:match}
Matching optical catalogs with each other depends on the cluster finding method used to extract a catalog, the kind of dataset used, the redshift range, and the identification of the cluster center. In this section we compare the $\mathtt{GalWCat19}$ catalog with previous cluster catalogs by matching them in a traditional way as performed in the literature (see e.g., \citealp{Wen12,Banerjee18}). This task is accomplished by searching within a given radius and velocity gap (or redshift) from each GalWeight cluster center. We adopt a search radius of 1.5 $ h^{-1}$Mpc ($\sim$ twice the mean value of $R_{200}$ in our catalog). Also, we adopt the velocity gap of $\pm1500$ km s$^{-1}$ ($\sim$ redshift difference of 0.01). We compare $\mathtt{GalWCat19}$ with previous catalogs, including Yoon \citep{Yoon08}, GMBCG \citep{Hao10}, WHL \citep{Wen12}, redMaPPer \citep{Rykoff14}, Tempel \citep{Tempel14}, and AMF \citep{Banerjee18} catalogs. 
{\bf
{Note that some catalogs provided both spectroscopic and photometric redshifts for clusters. In that case we match our catalog with each of these redshifts as shown in Table \ref{tab:Match}.
}}

The procedure used to compare $\mathtt{GalWCat19}$ with other catalogs is as follows. \\
\noindent 1. In an overlapping redshift range ($z_{over}$) between $\mathtt{GalWCat19}$ and the reference catalog we determine the number of clusters in $\mathtt{GalWCat19}$ ($Ngw$) and the corresponding number of clusters in the reference catalog ($Ncat$).\\ 
\noindent  2. We calculate how many clusters match ($Nmat$) in a radius of 1.5 $h^{-1}$ Mpc and velocity gap of $\pm1500$ km s$^{-1}$ relative to $\mathtt{GalWCat19}$ cluster center. \\
\noindent 3. We determine the number of clusters which are included in $\mathtt{GalWCat19}$ and are not identified  by the reference catalog ($Ngw_{o}$).\\
\noindent 4. We calculate the number of clusters which are not identified by $\mathtt{GalWCat19}$ but included in the reference catalog ($Ncat_{o}$). \\
\noindent 5.  We determine the number of clusters not identified  by $\mathtt{GalWCat19}$ but included in the reference catalog for which there are at least 8 galaxies in a projected distance of $R_{p} = 0.5~h^{-1}$ Mpc and velocity range $=\pm1500$ km s$^{-1}$ from the cluster center ($Ncat_{o,FG}$) (the cutoff condition of our catalog). \\
\noindent 6. Finally, the ratios $Rmat=Nmat/Ngw$, and $Rcat_{o,FG}=Ncat_{o,FG}/Ngw$ are calculated (see Table \ref{tab:Match}).

\begin{table*} \centering
\caption{Matches with other catalogs.}
\label{tab:Match}
\scriptsize
\begin{tabular}{ccccccccccccccc}\hline
&&&\multicolumn{6}{c}{numbers}&&\multicolumn{2}{c}{ratios}\\
Catalog&$z_{over}$&& 
$Ngw$& $Ncat$&$Nmat$&$Ngw_{o}$&$Ncat_{o}$&$Ncat_{o,FG}$&& 
$Rmat$&$Rcat_{o,FG}$&&z-type&reference\\
%&$Rgw_o$&$Rcat_o$
(1)&(2) &&(3)&(4)&(5)&(6)&(7)&(8)&&(9)&(10)&&(11)&(12)\\ 
\hline
Yoon        &0.049 : 0.101 &&950 &924 &417 &533 &507 &266 &&{\bf{0.439}} &0.280&&spect&\citet{Yoon08}\\
GMBCG       &0.099 : 0.196 &&650 &3677 &182 &468 &3495 &11 &&{\bf{0.280}} &0.017&&photo&\citet{Hao10}\\
GMBCG       &0.007 : 0.196 &&1800 &2993 &440 &1360 &2553 &97 &&{\bf{0.244}} &0.054&&spect&\citet{Hao10}\\
WHL        &0.049 : 0.196 &&1580 &15601 &726 &854 &14875 &82 &&{\bf{0.459}} &0.052&&photo&\citet{Wen12}\\
WHL        &0.043 : 0.196 &&1636 &9117 &907 &729 &8210 &294 &&{\bf{0.554}} &0.180&&spect&\citet{Wen12}\\
redMaPPer &0.079 : 0.196 &&1073 &2248 &429 &644 &1819 &25 &&{\bf{0.400}} &0.023&&photo&\citet{Rykoff14}\\
redMaPPer &0.050 : 0.196 &&1569 &1410 &381 &1188 &1029 &24 &&{\bf{0.243}} &0.015&&spect&\citet{Rykoff14}\\
Tempel   &0.007 : 0.196 &&1800 &3296 &1230 &570 &2066 &482 &&{\bf{0.683}} &0.268&&spect&\citet{Tempel14}\\
AMF            &0.044 : 0.196 &&1628 &7033 &848 &780 &6185 &184 &&{\bf{0.521}} &0.113&&photo&\citet{Banerjee18}\\
\hline
\end{tabular}
\begin{tablenotes}
\item
\noindent
Columns: (1) catalog name; (2) intersecting redshift range; (3) number of clusters in the redshift range of the catalog; (4)  number of clusters in the redshift range of the GalWeight catalog; (5) number of clusters that matches with GalWeight catalog; (6) number of clusters that is included in GalWeight catalog and missed by the other catalog; (7) number of clusters that is missed by GalWeight catalog and included in the other catalog; (8) same as column (7) but for clusters that only satisfy the cutoff condition of our catalog; (9-10) the ratios $Rmat=Nmat/Ngw$, and $Rcat_{o,FG}=Ncat_{o,FG}/Ngw$; (11) the redshift type; (12) the reference of the catalog. 
\end{tablenotes}
\end{table*} 

% Fig4___________________________________________________
\begin{figure*} \hspace*{-0.5cm}
\includegraphics[width=28cm]{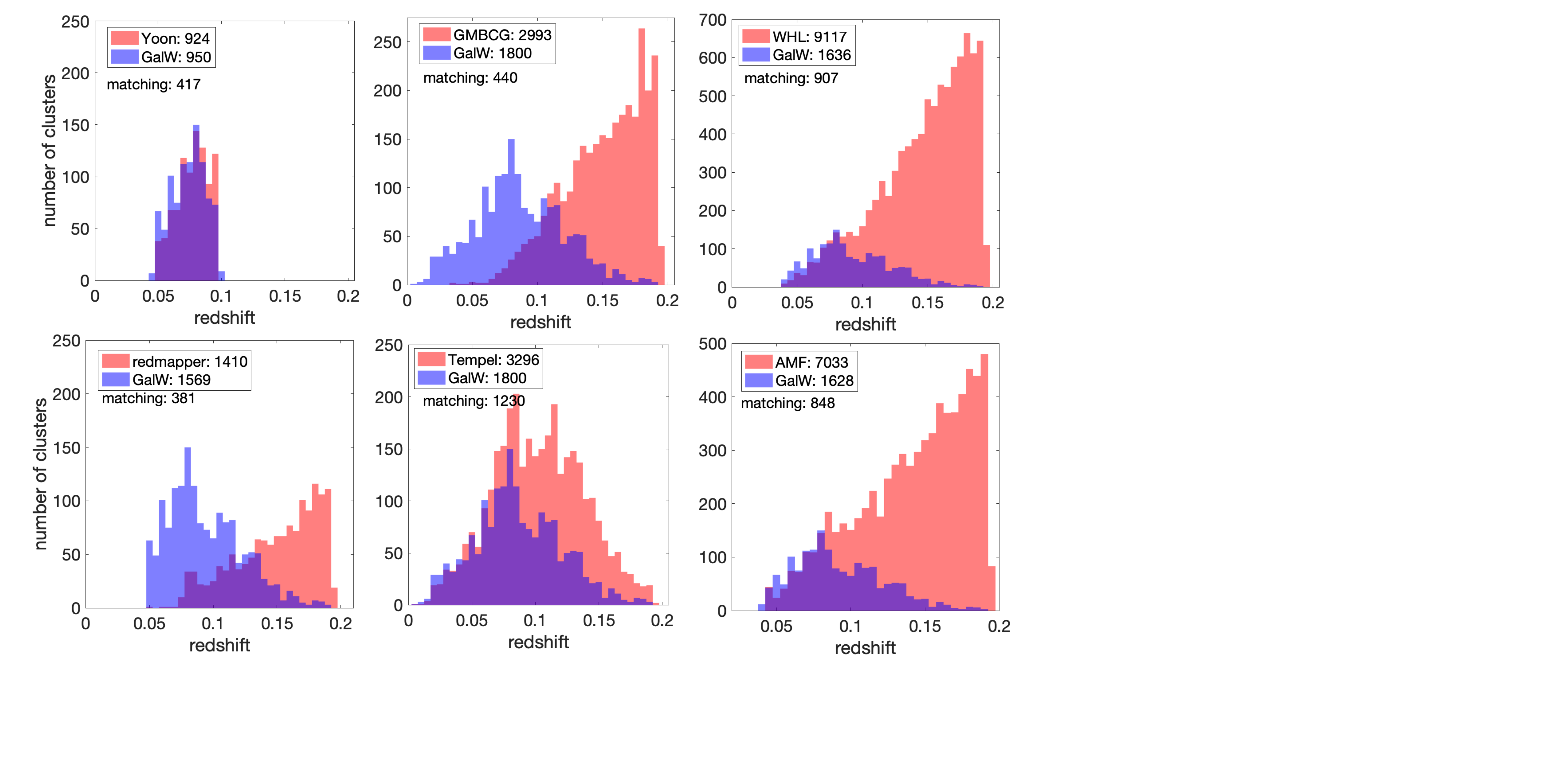} \vspace{-2.5cm}
\caption{Matching $\mathtt{GalWCat19}$ (blue histograms) with six optical catalogs (red histograms). The histograms of Yoon, GMBCG, WHL, redMaPPer, and Tempel are derived from spectroscopic redshifts provided by each catalog, while the histogram of AMF is derived from photometric redshift that does not provide spectroscopic data.}
\label{fig:match}
\end{figure*}

A summary of each catalog, cluster finding method, and redshift range is descried below. We refer the reader to the reference of each catalog for more details.\\
\noindent 1. The Yoon catalog:-\\
Yoon catalog is a local density cluster finder catalog \citep{Yoon08} applied on SDSS-DR5 using the spectroscopic and photometric redshift dataset. The catalog identified 924 clusters in a  spectroscopic redshift range of $z_{sp} = [0.049, 0.101]$. The number of matched clusters is 417 out of 950 $\mathtt{GalWCat19}$ clusters in the overlapping redshift range.

\noindent 2. The GMBCG catalog:-\\
GMBCG is a red-sequence plus brightest cluster galaxy cluster finder catalog \citep{Hao10} applied on SDSS-DR7 using the photometric redshift dataset. The catalog identified $\sim$ 50,000 clusters in a photometric redshift range of $z_{ph} = [0.1, 0.55]$. The catalog also provided spectroscopic redshift for 2,993 clusters in a range of  $z_{sp} = [0.007, 0.196]$. There are 440 matched clusters out of 1,800 in the overlapping spectroscopic redshift range.

\noindent 3. The WHL catalog:-\\
WHL is a red-sequence cluster finder catalog \citep{Wen12} applied on SDSS-DR8 using the photometric redshift ($z_{ph}$) dataset. The catalog identified 132,684 clusters in a photometric redshift range of $z_{ph} = [0.05, 0.785]$. The catalog provided spectroscopic redshift for 9,117 clusters in a range of $z_{sp} = [0.043, 0.196]$. The number of matched clusters is 912 out of 1695 in the overlapping spectroscopic redshift range.

\noindent 4. The redMaPPer catalog:-\\
redMaPPer is a red-sequence cluster finder catalog \citep{Rykoff14} applied on SDSS-DR8 using the photometric redshift dataset. The catalog identified 25,325 clusters in a photometric redshift range of $z_{ph} = [0.08, 0.55]$. The catalog also provided spectroscopic redshift for 1,410 clusters in a range of $z_{sp} = [0.050, 0.196]$. The number of matched clusters are 381 out of 1,569 in the overlapping spectroscopic redshift range.

\noindent 5. The Tempel catalog:-\\
Tempel catalog is based on a modified friends-of-friends method \citep{Tempel14}, and is applied on the spectroscopic sample of galaxies of SDSS-DR10. The catalog identified 82,458 clusters in a spectroscopic redshift range of $z_{sp} = [0.08, 0.2]$.  There are 3296 clusters in the catalog with masses $\geq 0.4 \times10^{14}~h^{-1}M_{\odot}$  (the cutoff mass of $\mathtt{GalWCat19}$) and number of galaxy members $= 4$ in $R_{200}$. The number of matched clusters is 1,230 out of 1800 in the spectroscopic overlapping redshift  range.

\noindent 6. The AMF catalog:-\\
AMF catalog \citep{Banerjee18} is based on an adaptive matched filter technique applied to SDSS-DR9. The catalog identified 46,479 galaxy clusters in a photometric redshift range of $z_{ph} = [0.045, 0.641]$. There are 7,033 clusters in the overlapping redshift $z_{ph} = [0.045, 0.196]$. The number of matched clusters is 848 out of 1,628 in the overlapping photometric redshift range.

As shown in Table \ref{tab:Match}, the matching rate, $Rmat=Nmat/Ngw$ varies from 0.24 to 0.68 depending on the cluster finding method used to extract a catalog, the dataset used, redshift range, and the identification of the cluster center. These are the main factors  that explain why the $\mathtt{GalWCat19}$ miss clusters relative to other catalogs and vice versa. Also, we expect that our catalog miss poor or low-mass clusters. This is because we cut the catalog at cluster masses of $M_{200} \geq 0.4 \times10^{14}~h^{-1}M_{\odot}$ and with the condition that the number of galaxies within a cylinder of $R_{p} = 0.5 h^{-1}$ Mpc and velocity range $=\pm1500$ km s$^{-1}$ is at least 8 galaxies. Moreover, for the catalogs extracted from photometric redshifts (GMBCG, WHL, redMaPPar, and AMF) the number of clusters at high redshift ($\sim 0.2$) is huge relative to $\mathtt{GalWCat19}$ which is extracted from spectroscopic redshifts. 
This is because the number of galaxies (and consequently the number of clusters) that have photometric redshifts is very large relative to the spectroscopic ones.

%################################################################################
\subsection{Velocity dispersion vs. Mass relation} \label{sec:VM}

% Fig5___________________________________________________
\begin{figure*} \hspace*{-0.5cm}
\includegraphics[width=21.5cm]{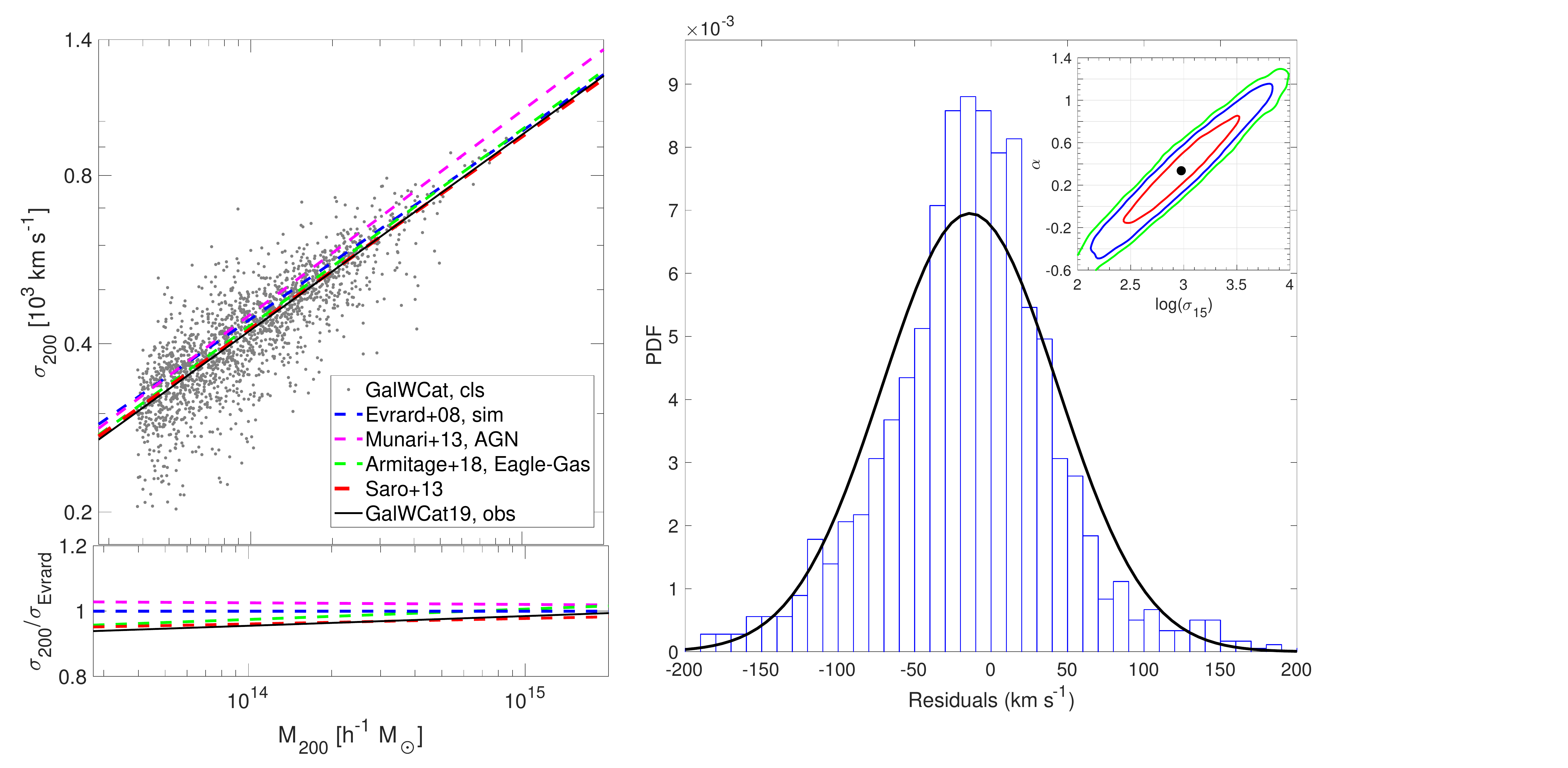} \vspace{-0.5cm}
\caption{Top left panel: Velocity dispersion $\sigma_{200}$ vs. virial mass $M_{200}$ for 1,800 clusters in the $\mathtt{GalWCat19}$ catalog. The gray points show the $\mathtt{GalWCat19}$ clusters and the solid black line represents the best-fit relation from Equation \ref{eq:Evr}. The blue, purple, green, and red dashed lines show the relations for \citet{Evrard08}, \citet{Munari13} \citet{Saro13}, and \citet{Armitage18} derived from cosmological simulations, respectively. As shown, the GalWeight relation matches the models remarkably well, indicating the accuracy of the GalWeight to constrain cluster membership, and consequently determine cluster masses. Bottom left panel: best-fit relations relative to the \citet{Evrard08} result. Right panel: The distribution of residual of velocity dispersion of clusters from the best-fit line,  along with best-fit the Gaussian curve. The inner right panel shows the best-fit parameters of Equation \ref{eq:Evr} with $1, 2 , 3 \sigma$ confidence intervals.}
\label{fig:sigmamass}
\end{figure*}

Estimating cluster masses accurately is a significant challenge in astronomy, since it is not a directly observable quantity. The use of velocity dispersion as a proxy for cluster mass has been shown to be particularly effective at low redshift compared to other techniques. \citet{Sereno15} showed that the intrinsic scatter in the $\sigma - M_{WL}$ relation was $\sim 14\%$ as opposed to $\sim 30\%$, $\sim 25\%$, and $\sim 40\%$ for X-ray luminosity, SZ flux, and optical richness, respectively. Also, since galaxies are nearly collisionless tracers of the gravitational potential, one expects velocity dispersion to be more robust than X-ray and SZ mass proxies.

\citet{Evrard08} (Evrard+08) found that the $\sigma - M$ relation for dark matter particles was close to the expected virial scaling relation of $\sigma \propto M^{1/3}$, with a minimal scatter of $\sim 5 \%$, and was insensitive to cosmological parameters. \citet{Munari13} (Munari+13), \citet{Saro13} (Saro+13), and \citet{Armitage18} (Armitage+18) investigated the $\sigma -M$ relation using hydrodynamical and semi-analytic simulations in order to understand how including baryonic physics in simulations affected the relation. Compared to the relation derived purely from N-body simulations (Evrard+08), the relations found by Munari+13, Saro+13 Armitage+18 suggested that galaxies introduce a bias in velocity relative to the DM particles (see Figure \ref{fig:sigmamass}). This bias can be either positive (a larger $\sigma$ for a given $M$ than what the DM particles have) or negative (a smaller $\sigma$ for a given $M$ than what the DM particles have), depending on the halo mass, redshift and physics implemented in the simulation (e.g., \citealp{Saro13,Old13,Wu13}). Also, Saro+13 concluded that the effect of the presence of interlopers on the estimated velocity dispersion could be the dominant source of uncertainty (up to $\sim49\%$).  However, the more sophisticated interloper rejection techniques, such as caustic \citet{Diaferio99} and GalWeight techniques \citet{Abdullah18} could result in a reduced uncertainty when calculating the velocity dispersion.

Following \citet{Evrard08}, the $\sigma_{200} - M_{200}$ relation can be expressed as 

\begin{equation} \label{eq:Evr}
\sigma_{200}=\sigma_{15}\left[\frac{h(z)~M_{200}}{10^{15}M_\odot}\right]^\alpha
\end{equation}

\noindent where $\sigma_{15}$ is the normalization at mass $10^{15}~h^{-1}M_{\odot}$, and $\alpha$ is the logarithmic slope. We follow \citet{Kelly07} and \citet{Mantz16b} to determine these two parameters in the log-log space of $\sigma_{200}$ and $M_{200}$.

The scatter, $\delta_{\log\sigma}$, in the $\sigma_{200} - M_{200}$ relation, defined as the standard deviation of $\log(\sigma)$ about the best-fit relation (see e.g., \citealp{Evrard08,Lau10}), is given by

\begin{equation} \label{eq:sct}
\delta_{\log\sigma} = \sqrt{\frac{1}{N} \sum_{i=1}^N \log(\sigma_i/\sigma_{fit})^2}
\end{equation}

\noindent where $\sigma_i$ is the velocity dispersion of the $i^{th}$ cluster and $\sigma_{fit}$ is the best-fit value. For $\sigma_{200}$ and $M_{200}$ determined by the virial mass estimator we get $\sigma_{15} = 946\pm52~ \mbox{km} ~ \mbox{s}^{-1}$, and $\alpha = 0.349\pm0.142$ with a scatter of $\delta_{\log\sigma} =0.06\pm0.04$ for all clusters with mass $M_{200} \geq 0.4 \times 10^{14}h^{-1}M_{\odot}$.

Figure \ref{fig:sigmamass} shows the $\sigma_{200}-M_{200}$ relation for the 1,800 clusters in the $\mathtt{GalWCat19}$ catalog. The gray points represent the $\mathtt{GalWCat19}$ clusters and the solid black line is the best-fit relation from Equation \ref{eq:Evr}. The blue, purple, green, and red dashed lines show the relations from Evrard+08, Munari+13, Saro+13, and Armitage+18 which were derived from cosmological simulations. Generally speaking, the $\mathtt{GalWCat19}$ line matches the models remarkably well, indicating the effectiveness of the GalWeight technique in constraining cluster membership, and consequently in determining cluster mass. However, we cannot make a quantitative comparison between the observed line and the other three models of Evrard+08, Munari+13 and Armitage+18. This is because Evrard+08 derived this relation for purely dark matter particles without taking into account the effect of baryons and it is well-known that galaxies are biased tracers of dark matter particles. Moreover, even though Munari+13 and Armitage+18 included baryonic physics, their relations were derived from the true members, while our sample is contaminated by interlopers (projection effects). The only relation that took into account the baryonic physics and the projection effect (i.e., presence of interlopers) is Saro+13. As shown in the Figure \ref{fig:sigmamass}, Saro+13 model is the closest to our observed line. 

Finally, we stress that the calculated velocity dispersion and consequently the cluster mass are scattered by the presence of interlopers as well as other factors which were discussed above in \S \ref{sec:recover}. In order to study this scaling relation in detail one should take into consideration all of these factors and utilize both hydrodynamical and semi-analytic models to digest the different sources of scatter and uncertainties. This is certainly out of the scope of this paper and we defer this investigation to a later paper.

\section{Conclusion} \label{sec:conc}
In this paper we used the SDSS-DR13 spectroscopic dataset to identify and analyze a catalog of 1,800 galaxy clusters ($\mathtt{GalWCat19}$). The cluster sample has a mass range of  $(0.40 - 14) \times 10^{14}$ $h^{-1}$ M$_{\odot}$ and a redshift range $0.01 \leq z \leq 0.2$ with a total of 34,471 galaxy members identified within the virial radii of the 1,800 clusters.

The clusters were identified by a simple algorithm that looks for the Finger-of-God effect (the distortion of the peculiar velocities of its core members along line-of-sight). The FOG effect was detected by assuming a cylinder of radius $R_{cy} = 0.5 h^{-1}$ Mpc ($\sim$ the width of FOG), and height $3000$ km s$^{-1}$ ($\sim$ the length of FOG) centered at each galaxy in our sample. We selected all overdensity regions with the condition that the cylinder has at least eight galaxies. The completeness of our sample identified by the FG algorithm, was tested by the Bolshoi simulation.  The completeness to identify locations of clusters with at least eight galaxies was approximately 100\% for clusters with masses $M_{200} > 2 \times 10^{14}~h^{-1}M_{\odot}$, while it dropped to $\approx 92\%$ for clusters with masses $M_{200} > 0.4 \times 10^{14}~h^{-1}M_{\odot}$.

The membership of each detected cluster was assigned by the GalWeight technique. Then, we used the virial theorem and NFW mass profile in order to determine dynamical parameters for each cluster from its galaxy members. This integrated procedure was applied to HOD2 and SAM2 mock catalogs recalled from \citealp{Old15} to test its efficiency in recovering cluster mass. GalWeight performs well in comparison to most other mass estimators described in \citealp{Old15} for both the HOD2 and SAM2 models. In particular, the rms differences of the recovered mass by GalWeight relative to the fiducial cluster mass are 0.26 and 0.28 for the HOD2 and SAM2, respectively. Furthermore, the rms error produced by GalWeight was among the lowest of all other methods that depend on the phase-space and velocity dispersion to calculate mass.

Using the virial mass estimator we determined the virial radius and its corresponding virial mass for each cluster. We then used NFW mass profile to determine the dynamical parameters of each cluster at density $\rho = \Delta  \rho_c$, for overdensities $\Delta = [500, 200, 100, 5.5]$. We assumed that the virial radius is at $\Delta$ = 200 and turnaround radius is at $\Delta$ = 5.5. We introduced a cluster catalog for the dynamical parameters derived by virial mass estimator and NFW model. The derived parameters for each cluster are radius, number of members, velocity dispersion and mass at different overdensities, plus the NFW parameters: scale radius, mass at scale radius, and concentration. We also introduced a membership catalog that correspond to the cluster catalog. The description of the catalogs are introduced in appendix \ref{app:catalogs}.

Finally, we showed that the cluster velocity dispersion scales with total mass for $\mathtt{GalWCat19}$ as 
$\log(\sigma_{200})=\log(946\pm52~ \mbox{km} ~ \mbox{s}^{-1}) +(0.349\pm0.142)\log\left[h(z) ~ M_{200}/10^{15}M_\odot\right]$ with scatter $\delta_{\log\sigma} = 0.06$. This relation was well-fitted with the theoretical relations derived from the N-body simulations. 

\section*{FUTURE WORK}
In future work, we aim to: (i) study the halo-mass, stellar mass, and luminosity functions of $\mathtt{GalWCat19}$ to constrain the matter density of the universe, $\Omega_m$, and the normalization of the linear power spectrum, $\sigma_8$; (ii) investigate the stellar mass and luminosity function of member galaxies of their hosting clusters; (iii) study the shape of velocity dispersion profiles of $\mathtt{GalWCat19}$ and compare with Multi-dark simulations in order to recover cluster mass. (iv) study the connection between stellar mass (or luminosity) and dark matter halo; (v) investigate the effect of environment on the properties of member galaxies such as size, and quenching of star formation and segregation of star forming and quiescent galaxies on a small scale; (vi) investigate the adaptation of the GalWeight technique to recover cluster mass and cluster mass profile; (vii)) study the correlation function of galaxy clusters and the signature of Acoustic Baryonic Oscillation (BAO) to constrain cosmological parameters using the $\mathtt{GalWCat19}$.

\section*{ACKNOWLEDGMENTS}
We thank Gary Mamon for the useful discussion to fit NFW model using maximum likelihood technique. We also thank 
Steven Murray for the publicly available code \verb|HMFcalc| and his guideline to run the code. We also thank Stefano Andreon for his useful comments. Finally, we appreciate the comments and suggestions of the reviewer which improved this paper.
This work is supported by the National Science Foundation through grant AST-1517863, by HST program number GO-15294,  and by grant number 80NSSC17K0019 issued through the NASA Astrophysics Data Analysis Program (ADAP). Support for program number GO-15294 was provided by NASA through a grant from the Space Telescope Science Institute, which is operated by the Association of Universities for Research in Astronomy, Incorporated, under NASA contract NAS5-26555. Lyndsay Old acknowledges the European Space Agency (ESA) Research Fellowship.

\begin{appendices}

\section{Description of the Catalogs in the $\mathtt{GalWCat19}$ release} \label{app:catalogs}
The $\mathtt{GalWCat19}$ release consists of two catalogs. The first catalog lists the coordinates and the dynamical parameters of each galaxy cluster. The second catalog lists the coordinates of the member galaxies belonging to each cluster. The two catalogs are publicly-available at the website\footnote{\url{https://mohamed-elhashash-94.webself.net/galwcat}}.

\subsection{Description of the Cluster Catalog}
The cluster catalog contains the following information (column numbers are given in square brackets):\\ 

$[1]$\, \texttt{clsid} -- our unique identification number for clusters;\\
$[2-3]$\, \texttt{raj2000, dej2000} -- right ascension and declination of the cluster center in deg;\\
$[4]$\, \texttt{$z_{cls}$} -- cluster redshift, calculated as an average over all cluster members;\\ 
$[5]$\, \texttt{$v_{cls}$} -- radial velocity of the cluster in units of \mbox{km~s$^{-1}$};\\
$[6]$\, \texttt{$D_{cls}$} -- comoving distance of the cluster in units of $h^{-1}\,$Mpc;\\
$[7]$\, \texttt{$R_{500}$} --  the radius from the cluster center at which the density $\rho=\Delta_{500}\rho_c$ in units of $h^{-1}\,$Mpc;\\
$[8]$\, \texttt{$N_{500}$} -- number of members of the cluster within $R_{500}$;\\
$[9]$\, \texttt{$\sigma_{500}$} --  velocity dispersion in \mbox{km~s$^{-1}$} of the cluster within $R_{500}$;\\
$[10-11]$\, \texttt{$\sigma\_Err(-)_{500}$, $\sigma\_Err(+)_{500}$} --  lower and upper errors of $\sigma_{500}$ in \mbox{km~s$^{-1}$}, obtained via 1000 bootstrap resampling;\\
$[12]$\, \texttt{$M_{500}$} -- mass of the cluster at $R_{500}$ in units of $10^{14}~h^{-1}M_\odot$;\\
$[13]$\, \texttt{$M\_Err_{500}$} -- error in $M_{500}$ in units of $10^{14}~h^{-1}M_\odot$;\\
$[14]$\, \texttt{$R_{200}$} --  the radius from the cluster center at which the density $\rho=\Delta_{200}\rho_c$ in units of $h^{-1}\,$Mpc;\\
$[15]$\, \texttt{$N_{200}$} -- number of members of the cluster within $R_{200}$;\\
$[16]$\, \texttt{$\sigma_{200}$} -- velocity dispersion in \mbox{km~s$^{-1}$} of the cluster within $R_{200}$;\\
$[17-18]$\, \texttt{$\sigma\_Err(-)_{200}$, $\sigma\_Err(+)_{200}$} --  lower and upper error of $\sigma_{200}$ in \mbox{km~s$^{-1}$}, obtained via 1000 bootstrap resampling;\\
$[19]$\, \texttt{$M_{200}$} -- mass of the cluster at $R_{200}$ in units of $10^{14}~h^{-1}M_\odot$;\\
$[20]$\, \texttt{$M\_Err_{200}$} -- error in $M_{200}$ in units of $10^{14}~h^{-1}M_\odot$;\\
$[21]$\, \texttt{$R_{100}$} --  the radius from the cluster center at which the density $\rho=\Delta_{100}\rho_c$ in units of $h^{-1}\,$Mpc;\\
$[22]$\, \texttt{$N_{100}$} -- number of members of the cluster within $R_{100}$;\\
$[23]$\, \texttt{$\sigma_{100}$} --  velocity dispersion in \mbox{km~s$^{-1}$} of the cluster within $R_{100}$;\\
$[24-25]$\, \texttt{$\sigma\_Err(-)_{100}$, $\sigma\_Err(+)_{100}$} --  lower and upper errors of $\sigma_{100}$ in \mbox{km~s$^{-1}$}, obtained via 1000 bootstrap resampling;\\
$[26]$\, \texttt{$M_{100}$} -- mass of the cluster at $R_{100}$ in units of $10^{14}~h^{-1}M_\odot$;\\
$[27]$\, \texttt{$M\_Err_{100}$} -- error in $M_{100}$ in units of $10^{14}~h^{-1}M_\odot$;
$[28]$\, \texttt{$R_{5.5}$} --  the radius from the cluster center at which the density $\rho=\Delta_{5.5}\rho_c$ in units of $h^{-1}\,$Mpc;\\
$[29]$\, \texttt{$N_{5.5}$} -- number of members of the cluster within $R_{5.5}$;\\
$[30]$\, \texttt{$\sigma_{5.5}$} --  velocity dispersion in \mbox{km~s$^{-1}$} of the cluster within $R_{5.5}$;\\
$[31-32]$\, \texttt{$\sigma\_Err(-)_{5.5}$, $\sigma\_Err(+)_{5.5}$} --  lower and upper errors of $\sigma_{5.5}$ in \mbox{km~s$^{-1}$}, obtained via 1000 bootstrap resampling;\\
$[33]$\, \texttt{$M_{5.5}$} -- mass of the cluster at $R_{5.5}$ in units of $10^{14}~h^{-1}M_\odot$;\\
$[34]$\, \texttt{$M\_Err_{5.5}$} -- error in $M_{5.5}$ in units of $10^{14}~h^{-1}M_\odot$;\\
$[35]$\, \texttt{$R_s$} --  scale radius of NFW model  in units of $h^{-1}\,$Mpc;\\
$[36]$\, \texttt{$R_s\_Err$} -- error in  scale radius of NFW model  in units of $h^{-1}\,$Mpc;\\
$[37]$\, \texttt{$M_s$} -- scale mass of the cluster at $R_{s}$ in units of $10^{14}~h^{-1}M_\odot$;\\
$[38]$\, \texttt{$M_s\_Err$} -- error in $M_{s}$ in units of $10^{14}~h^{-1}M_\odot$;\\
$[39]$\, \texttt{$c$} --  cluster concentration of NFW model

\subsection{Description of the Galaxy Catalog}
The catalog of the member galaxies correspond to the cluster catalog:\\
$[1]$\, \texttt{clsid} -- our unique identification number for clusters that member galaxies belong to;\\
$[2-3]$\, \texttt{raj2000, dej2000} -- right ascension and declination of the galaxy in deg;\\
$[4]$\, \texttt{$z_{g}$} -- observed redshift of the galaxy as given in the SDSS-DR-13;\\ 

\end{appendices}

\bibliography{ref1}

\end{document}